\documentclass[preprintnumbers,amsmath,amssymb]{revtex4}
\usepackage{graphicx,color}
\usepackage{amsmath,amssymb,latexsym}

\begin{document}

\title{Conduction at the onset of chaos}

\author{Fulvio Baldovin}
\email{baldovin@pd.infn.it}
\affiliation{
Dipartimento di Fisica, Sezione INFN, CNISM, and
Universit\`a di Padova,
 Via Marzolo 8, I-35131 Padova, Italy
}

\begin{abstract}
After a general discussion of the thermodynamics of conductive
processes, we introduce specific observables enabling the connection
of the diffusive transport properties with the microscopic dynamics. 
We solve the case of Brownian particles, both
analytically and numerically, and address then whether aspects of the
classic Onsager's picture generalize to the non-local non-reversible
dynamics described by logistic map iterates. 
While in the chaotic case numerical evidence of a monotonic
relaxation is found, at the onset of chaos complex relaxation patterns
emerge. 
\end{abstract}

\maketitle

\section{Introduction}
\label{sec_intro}
Matter properties can be transported by convection or
conduction~\cite{hertel_01}. While the former is related to the flow of the center of
mass of the ``material points'' in which the system under study may be
decomposed, the latter is caused by
interactions between neighboring particles. 
In many theoretical approaches, microscopic interactions can be
effectively represented at a coarse-grained level and conductive properties 
related to few thermodynamic parameters characterizing the system. 
On this basis, Onsager~\cite{onsager_01} has been able to
understand within a general framework how a slightly perturbed system 
relaxes (or regresses) to equilibrium, linking
its microscopic or mesoscopic transport properties to a
(nonequilibrium) thermodynamic description. 
In view of the universal character of the Onsager approach, 
from a fundamental perspective it
becomes particularly interesting 
trying to understand whether parts of such description 
may also apply to domains in which the
underlying dynamics is not Hamiltonian, and, e.g., microscopic reversibility
is lost. 
Nonlinear maps of the logistic class are a specific and simple enough
setting where this research perspective can be tested,
also thanks to the fact that a number of statistical mechanics techniques
have been successfully designed to their analysis~\cite{schuster_01,beck_01,robledo_01}.

The present study is a first effort in this direction. 
We offer a context in which the 
thermodynamics of conduction can be directly related to simple dynamical observables.
Such a context is analytically and numerically worked out for Brownian
particles, and explicit contact is established among the
thermodynamic relaxation properties, the system geometry, and
dynamical coefficients.
Finally, the Brownian dynamics is replaced by logistic map
iterations, and numerical studies are performed both in the fully
chaotic case and at the onset of chaos. 
We find that while results for chaotic dynamics display a monotonic
regression to equilibrium, at the onset of chaos
an involved oscillatory behavior emerges.

\begin{figure}
\begin{center}
\includegraphics[width=.7\textwidth]{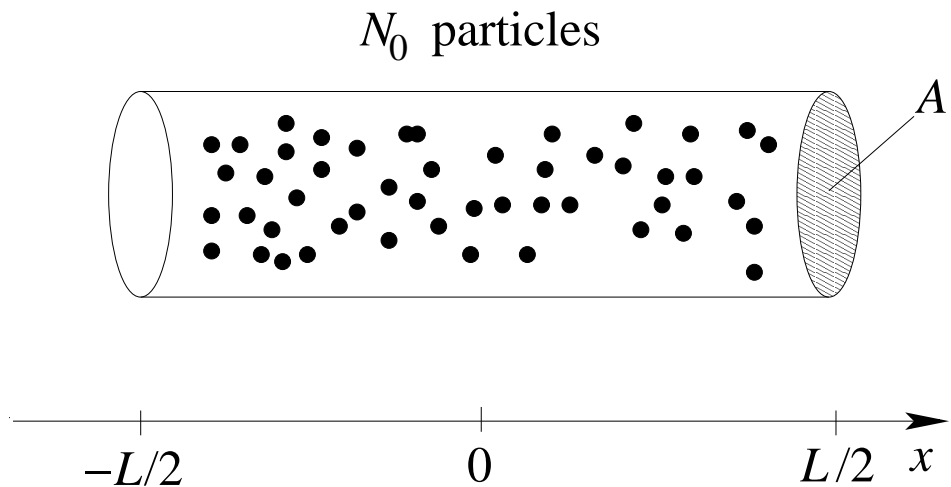} 
\caption{$N_0$ particles distributed within a cylinder of section $A$
  and length $L$.}
\label{fig_channel}
\end{center}
\end{figure}

\section{Thermodynamics of conduction}
In this Section, a general account of the thermodynamics of conduction
is given. We choose the language of continuum physics, where both
extensive and associated intensive thermodynamic variables becomes
local fields. Although, in view of the applications that follow, we
specialize our discussion to diffusion, analogous considerations
directly translate to other conduction processes, like e.g. the electric
or thermal ones. 
%After a general discussion applicable to small
%perturbations, we further assume that intensive parameters are
%linearly varying and discuss within this context the Onsager picture
%of the regression dynamics.

\subsection{General discussion}
Consider a system of $N_0$ equal particles confined within a cylinder of section $A$ and
radius much smaller than the length $L$ (see Fig.~\ref{fig_channel}). 
The cylinder is thermally, mechanically, and chemically isolated. 
In order to simplify the notations, particles are assumed to be
uniformly distributed in any cross-section of the cylinder, 
so that we are basically reduced to
a one-dimensional ($1d$) problem.  
Indicating as $N(x,t)$ the $3d$ particle distribution function, we will
be concerned below about its $n$th-order moments:
\begin{equation}
N_n\equiv A\;\int \mathrm{d} x\;x^n\;N(x,t)\,
\end{equation}
(if not otherwise indicated, integrals over $x$ are intended to span
the interval $[-L/2,L/2]$). 
Assuming local equilibrium, we may define a local entropy density
$\sigma(N(x,t))$. The system's (constrained) entropy becomes thus the functional
\begin{equation}
S[N]=
A\int \mathrm{d}x\;\sigma(N(x,t)).
\end{equation}
A measure about how far the system is from equilibrium is given by the first variation
\begin{equation}
\delta S[N]
=
-A\int \mathrm{d}x\;\frac{\mu}{T}(N(x,t))\;\delta N(x,t),
\end{equation}
where 
\begin{equation}
\frac{\mu}{T}(N(x,t))\equiv-\frac{\delta S[N]}{\delta N(x)}= -\frac{\partial \sigma(N(x,t))}{\partial N(x,t)}
\end{equation}
is the local intensive parameter associated to the
number of particles in the entropy representation of thermodynamics
\cite{callen_01}, 
and the variations of the density profile 
$\delta N(x,t)$ must satisfy the
impermeable-walls boundary condition 
$
\delta N_0=
A\;\int \mathrm{d} x\;\delta N(x,t)
=0.
$
While the equilibrium distribution $N_{\rm eq}(x)$ 
could be non-uniform, the extremal
requirement of zero first-variation $\delta S[N_{\rm eq}]=0$ implies that at equilibrium
$\mu/T$ is the same in any position $x$, as one may verify considering
the specific variation reallocating a particle from $x_1$ to $x_2$, 
$\delta N(x)=[\delta(x-x_2)-\delta(x-x_1)]/A$. 

The so-called fluctuation approximation amounts to a Taylor expansion
to quadratic order of $S$ around $N_{\rm eq}$:
\begin{equation}
S[N]
\simeq S[N_{\rm eq}]
+\frac{A}{2}\int \mathrm{d}x\;\left.\frac{\partial^2\sigma}{\partial N^2}\right|_{N_{\rm eq}(x)}\;\delta N(x,t)^2.
\label{eq_taylor_func}
\end{equation}
%in which the entropy can be equivalently
%regarded as a functional of the fluctuation $\delta N$ itself. 
The expression of the functional derivative as
\begin{equation}
\frac{\delta S[N]}{\delta N(x)}
=
%A\;
\left.\frac{\partial^2\sigma}{\partial N^2}\right|_{N_{\rm eq}(x)}\;\delta N(x,t)
\label{eq_delta_N_1}
\end{equation}
manifests the (local) generalized force -- linear in the fluctuation $\delta N$ 
with intensity regulated by the  thermodynamic response
$\partial^2\sigma/\partial N^2$ -- which drives the system
back to equilibrium. 
According to Einstein's formula~\cite{einstein_01,mauri_01}, the equilibrium probability for a fluctuation
is proportional to the exponential of the constrained
entropy, implying
\begin{equation}
\mathbb E\left[\delta N(x)^2\right]
=\int\frac{\mathcal{D}(\delta N(x))}{\mathcal{N}}\;\delta N(x)^2
\exp\left[
\frac{A}{2k_B}\int \mathrm{d}x\;\left.\frac{\partial^2\sigma}{\partial N^2}\right|_{N_{\rm eq}(x)}\;\delta N(x)^2
\right].
\end{equation}
A Gaussian integration~\cite{kardar_01} thus shows that
such response is directly linked to the average squared local fluctuation:
whence
\begin{equation}
\mathbb E\left[\delta N(x)^2\right]
=-\left[\frac{A\,L}{k_B}\left.\frac{\partial^2\sigma}{\partial N^2}\right|_{N_{\rm eq}(x)}\right]^{-1}.
\label{eq_delta_N_2}
\end{equation}

\subsection{Linearly varying intensive parameter}
If a system is sufficiently close to equilibrium, even with possibly
rough density profiles $N_{\rm eq}(x)$ the intensive
parameter $\mu/T$ can be assumed to be spatially smooth (see previous
Section).
In those cases in which $\mu/T$ is linearly varying along $x$,  a number of the above general
derivations assume a more transparent meaning.

Since in principle the distribution $N(x,t)$ can be characterized in
terms of all its moments $N_n$, we may think of the entropy functional
as a simple function of such moments~\cite{attard_01}:
\begin{equation}
S[N]\equiv S(N_0,N_1,\ldots).
\end{equation}
Retaining only the zero- and first-order moments in the right-hand
side and taking the functional derivative, we obtain the equation
\begin{eqnarray}
\;\frac{\delta S[N]}{\delta N(x)}
&=&\frac{\partial S(N_0,N_1)}{\partial N_0}
\;\frac{\delta N_0}{\delta N(x)}
+\frac{\partial S(N_0,N_1)}{\partial N_1}
\;\frac{\delta N_1}{\delta N(x)},
\nonumber\\
-\frac{\mu}{T}(x,t)
&=&
-\overline{\mu/T}
+\frac{\partial S(N_0,N_1)}{\partial N_1}
\;x,
\end{eqnarray}
where $\mu/T$ has been regarded directly as a function of $(x,t)$, 
and $\overline{\mu/T}=\mu/T(0,t)$ is the global intensive parameter. 
The latter result implies
\begin{equation}
\frac{\partial S(N_0,N_1)}{\partial N_1}
=-\nabla_x\frac{\mu}{T}(0,t),
\end{equation}
with  $\nabla_x(\mu/T)$ uniform within this
approximation.
One thus recognizes that 
the thermodynamic force restoring the 
fluctuation to equilibrium is now seen as
the gradient in the intensive parameter
$\mu/T$. 
Since, in view of the impermeable boundaries, $N_0$ is the same for any
distribution, it is not a relevant thermodynamic parameter and may be
safely neglected in what follows.
Indicating as $N_{1,{\rm eq}}$ the first moment of $N_{\rm eq}$,
the fluctuation approximation can now be written as 
\begin{equation}
S(N_1)
\simeq S(N_{1,{\rm eq}})
+\frac{1}{2}
\left.\frac{\partial^2 S}{\partial N_1^2}\right|_{N_{1,{\rm eq}}}
\;\delta N_1^2.
\end{equation}
The use of the Einstein's formula~\cite{einstein_01,mauri_01} for the probability of a
fluctuation $\delta N_1$ gives, for the generalized force,
\begin{equation}
-\nabla_x\frac{\mu}{T}(0,t)
\simeq 
\left.\frac{\partial^2 S}{\partial N_1^2}\right|_{N_{1,{\rm eq}}}
\;\delta N_1(t)
=-\frac{k_B}{\mathbb{E}[\delta N_1^2]}
\;\delta N_1(t).
\end{equation}

The time derivative of $\delta N_1$ is  closely related to the number
flux $J_N$. Take, for simplicity, a quasi-stationary state within
the cylinder, i.e., a situation in which the thermodynamic parameters
are almost time-independent. 
A quasi-stationary state can only be supported by the existence of 
uniform fluxes. In such a way, the number of particles entering
arbitrary small volumes in a given time interval is equal to
those  leaving it. 
Assuming thus $J_N(x,t)=\overline J(t)$ 
for $x\in[-L/2,L/2]$ (slowly varying in $t$), and 
$J_N(x,t)=0$ for $x\notin[-L/2,L/2]$, we have 
\begin{equation}
\nabla_x J_N(x,t)
=\overline{J}_N(t)\,\left[\delta(x+L/2)-\delta(x-L/2)\right].
\end{equation}
Plugging this result in the continuity equation, 
\begin{equation}
\partial_t\,\delta N(x,t)
=\partial_t\,N(x,t)=-\nabla_x J_N(x,t),
\end{equation}
we indeed obtain
\begin{equation}
\partial_t\,\delta N_1(t)=AL\;\overline J_N(t).
\end{equation}

\subsection{Onsager regression dynamics}
Consider a small fluctuation at time $t_0$, which can be monitored
through the first moment of the density profile, $\delta N_1(t_0)$.
According to Onsager~\cite{onsager_01}, the thermodynamic force determining
the behavior of $\delta N_1$ does not depend on whether the fluctuation is
spontaneous or generated by the application of an external field or
reservoir. 
It is possible to prove~\cite{attard_02} that the most
likely small-time behavior of $\delta N_1(t_0+\tau)$, \mbox{$\overline{\delta N_1}(t_0+\tau)$}, 
is linear both in the thermodynamic
force and in time:
\begin{equation}
\overline{\delta N_1}(t_0+\tau)
\;\begin{subarray}{c} 
\simeq\\
|\tau|\ll1
\end{subarray}\;
\delta N_1(t_0)-|\tau|\;\frac{\Lambda}{2}\;\nabla_x\frac{\mu}{T}(0,t_0),
\label{eq_onsager_1}
\end{equation} 
where $\Lambda$ is a (positive) coefficient encoding the transport
properties of the system (see below). 
Eq.~(\ref{eq_onsager_1}) applies to $|\tau|$ larger than the microscopic
(molecular) time-scale of the dynamics, but still small with respect
to the significant macroscopic evolution of the system~\cite{attard_02}.  
In terms of the number flux, Eq.~(\ref{eq_onsager_1}) can be rewritten
as
\begin{equation}
\overline J_N(t_0+\tau)
\simeq
\frac{1}{AL}
\;\frac{\overline{\delta N_1}(t_0+\tau)-\delta N_1(t_0)}{|\tau|}
=-\frac{\Lambda}{2AL}\;\nabla_x\frac{\mu}{T}(0,t_0).
\end{equation}
If the intensive parameter $\mu/T$ can be
split into chemical potential $\mu$ and temperature $T$, and the
latter can be assumed to be uniform along the system, this result is often
written as the {\it first Fick's law}~\cite{fick_01}:
\begin{equation}
\overline J_N
=-D\;\nabla_x N,
\end{equation}
with
\begin{equation}
D\equiv\frac{\Lambda}{2AL}
\;\frac{1}{T}
\;\frac{\partial\mu}{\partial N}
\end{equation}
the diffusion coefficient.

More generally, the coefficient $\Lambda$ is related to the fluctuation's
autocorrelation by the Green-Kubo relation \cite{green_01,kubo_01,kubo_02}:
\begin{equation}
\Lambda=
-\frac{2}{k_B\,|\tau|} 
\left(
\mathbb{E}[\delta N_1(t_0+\tau)\;\delta N_1(t_0)]
-\mathbb{E}\left[\delta N_1(t_0)^2\right]
\right),
\end{equation}
where $\mathbb{E}\left[\delta N_1(t_0)^2\right]=\mathbb{E}\left[\delta N_1^2\right]$
is an average over the equilibrium distribution. 
Equivalently, the 
system can be characterized in terms of a coefficient $\lambda$ which
singles out the dynamical part of the response and is defined as
\begin{equation}
\lambda\equiv 
\frac{k_B\;\Lambda}{\mathbb{E}[\delta N_1^2]}
=
-\frac{2}{|\tau|} 
\frac{
\mathbb{E}[\delta N_1(t_0+\tau)\;\delta N_1(t_0)]
-\mathbb{E}\left[\delta N_1(t_0)^2\right]
}{
\mathbb{E}\left[\delta N_1(t_0)^2\right]
}.
\end{equation}
In the case of simple diffusion, in the next Section we will
explicitly show how
$\lambda$ is related to the local transport coefficient $D$ and to the
global geometry of the system.
In terms of $\lambda$, Eq.~(\ref{eq_onsager_1})
recasts into
\begin{equation}
\overline{\delta N_1}(t_0+\tau)
\;\begin{subarray}{c} 
\simeq\\
|\tau|\ll1
\end{subarray}\;
\delta N_1(t_0)-\frac{\lambda\;|\tau|}{2}
\;\delta N_1(t_0).
\label{eq_onsager_2}
\end{equation} 
At larger $\tau$, for dynamical evolutions both Gaussian and Markovian the Doob's
theorem~\cite{doob_01,doob_02} 
ensures an exponential decay of the fluctuation given by
\begin{equation}
\overline{\delta N_1}(t_0+\tau)
=\delta N_1(t_0)
\;\mathrm{e}^{
-\frac{\lambda\,|\tau|}{2}
}.
\label{eq_doob}
\end{equation} 

In summary, 
we can appreciate that the nonequilibrium behavior of a
macroscopic observable can be synthesized in terms of a static response
coefficient $\mathbb{E}\left[\delta N_1^2\right]$ determining the strength of the
force restoring equilibrium, and of a dynamic response coefficient $\lambda$
describing the time decay of the nonequilibrium fluctuation. 
Conversely, by monitoring the time evolution of $\overline{\delta N_1}(t)$ sensible
information about $\lambda$ can be obtained.

\section{Conduction and Brownian motion}
One of the easiest setup in which the previous general nonequilibrium
discussion can be tested is perhaps that in which particles are
endowed with a Brownian dynamics. 
The typical situation within this context corresponds to the
interaction of the $N_0$ particles with an heat bath of smaller ones
(e.g., water) at a given temperature. 
Although the  underlying dynamics is assumed to be Hamiltonian, 
the interaction with the heat bath may be effectively represented by a
stochastic term, so that the heat bath particles are not explicitly
traced. Implicitly, the motion of the heat bath particles is assumed to compensate that of the
Brownian ones,  in order to preserve energy and momenta, and to be in
conditions of zero convection. 
In the overdamped regime~\cite{kardar_01}, the equation of motion for the
coordinate $x_i$ of each Brownian particle is given by the Langevin stochastic
differential equation
\begin{equation}
x_i(t+\mathrm{d}t)=x_i(t)+\sqrt{2D}\;\mathrm{d}W(t),
\label{eq_wiener}
\end{equation}
where $W(t)$ is a Wiener process~\cite{doob_02}, and reflecting boundary
conditions are applied as $x_i=\pm L/2$. In Physics' literature,
Eq.~\eqref{eq_wiener} corresponds to a Gaussian white noise
evolution for $\mathrm{d}x_i/\mathrm{d}t$~\cite{kardar_01}.

\begin{figure}
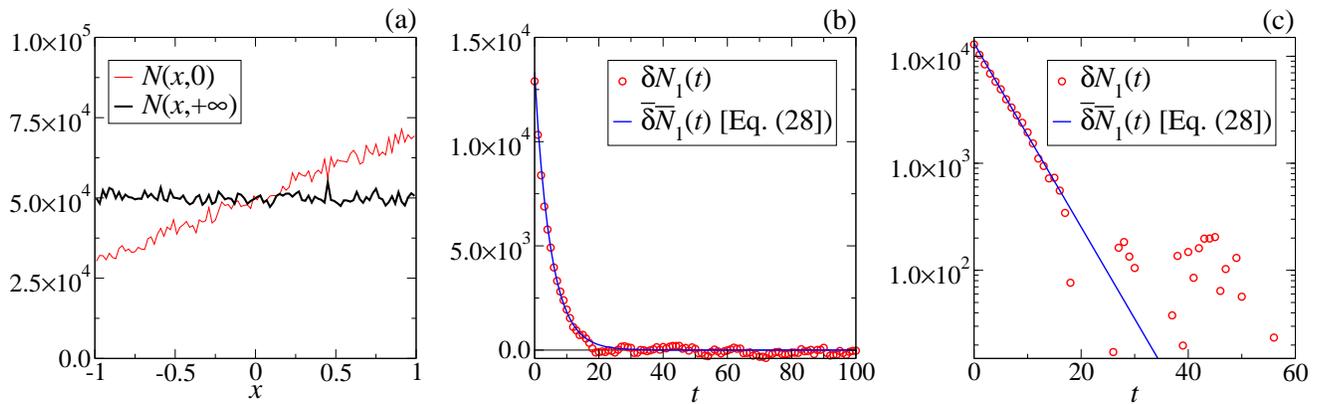

\begin{center}
\includegraphics[width=0.32\textwidth]{wiener_weight.eps}
\includegraphics[width=0.32\textwidth]{wiener_moment.eps}
\includegraphics[width=0.32\textwidth]{wiener_moment_log.eps}
\caption{
  Conduction and Brownian motion (simulation dimensionless units). 
  Eqs.~\eqref{eq_wiener} are numerically integrated for $N_0=10^5$
  independent particles  with
  reflecting boundaries at $\pm1$ ($L=2$) and $D=0.08$. 
  To calculate $N(x,t)$, the interval
  $[-1,1]$ is coarse grained in $100$ cells; $N(x,0)\equiv
  (1+a\,L\,x)\,N_0/(AL)$, with $x$ indicating the center of the cells
  and $a=0.2$. 
  (a): Initial and long-term distribution.
  (b,c): Numerical evaluation of $\delta N_1(t)$ (circles) is
  compared with Eq.~\eqref{eq_wiener_an} (line).
}
\label{fig_wiener}
\end{center}
\end{figure}

On the basis of the (Lagrangian) particles coordinates $x_i$, the
distribution function $N(x,t)$ is defined as
\begin{equation}
N(x,t)
\equiv \frac{1}{A}\;\sum_{i=1}^N\delta(x-x_i(t)),
\label{eq_dist}
\end{equation}
where $x$ is regarded as an Eulerian coordinate.
In the present case, there are two sources of
randomness for $N(x,t)$: one is the distribution of the initial
conditions $\{x_i(t_0)\}$; the other is because the dynamics itself is a
random process. 
As a consequence of the latter, the most likely time evolution of the distribution
function, $\overline{N}(x,t)$, satisfies the Fokker-Planck equation~\cite{kardar_01} 
\begin{equation}
\partial_t \overline{N}(x,t)
=D\;\nabla^2_x\overline{N}(x,t).
\label{eq_fp}
\end{equation}  
The solution is obtained by applying the appropriate Green function
for reflecting boundaries at $x=\pm L/2$ to the 
initial distribution $N(x_0,t_0)$~\cite{gardiner_01}:
\begin{equation}
\overline{N}(x,t)
=
\int\frac{\mathrm{d}x_0}{L}
\left[
1+2\sum_{n=1}^\infty
{\rm e}^{-\frac{n^2\pi^2\,D\,(t-t_0)}{L^2}}
\cos\left(n\,\pi\,\frac{2x+L}{2L}\right)
\cos\left(n\,\pi\,\frac{2x_0+L}{2L}\right)
\right]
N(x_0,t_0).
\end{equation}
Independently of $N_0$, the equilibrium distribution turns out to be uniform: 
$N_{\rm eq}(x)=\lim_{t\to+\infty}\overline{N}(x,t)=N_0/AL$,
and a straightforward calculation yields
\begin{equation}
\overline{\delta N_1}(t)
=-4AL
\sum_{
\begin{subarray}{c} 
n=1\\
n\;\textrm{odd}
\end{subarray}
}^{+\infty}
\frac{{\rm e}^{-\frac{n^2\pi^2\,D\,(t-t_0)}{L^2}}}{n^2\pi^2}
\int\mathrm{d}x_0
\;\cos\left(n\,\pi\,\frac{2x_0+L}{2L}\right)
\;N(x_0,t_0).
\label{eq_wiener_an}
\end{equation}
If $N(x_0,t_0)$ is sufficiently close to $N_{\rm eq}$,
Eq.~\eqref{eq_wiener_an} is dominated by the $n=1$ term, and we recover 
Eq.~\eqref{eq_doob} with 
\begin{equation}
\lambda=\frac{2\pi^2\,D}{L^2}.
\end{equation} 
As anticipated, we thus see that $\lambda$ is affected by 
both local transport properties and
global aspects of the geometry of the system.
In Fig.~\ref{fig_wiener} the numerical simulation of a system of
particles described by Eq.~\eqref{eq_wiener} is compared with the
analytical results. 

\section{Conduction and chaotic dynamics}
In this and in the following Section we address the phenomenology of
the regression of a nonequilibrium distribution within the context of
the logistic map.
The basic question we would like to explore is whether some of the
general Onsager results do generalize to such a dynamics.  
Before entering into details, some words of caution are in order. 
In the case of the logistic map, basic assumptions ordinarily
underlying the Onsager discussion are posed into question. 
First, the logistic map's dynamics is not local: being conceptually
the result of a Poincar\'e section on an orbit, for the logistic map ``$t$'' becomes a
discrete iteration time and at $t+1$ the iterates are mapped to a space
location typically far from that occupied at $t$. Second, the dynamics is
inherently non-reversible: the preimage of each iterate at time $t$
corresponds to two distinct points.
In view of these remarks, the study of the regression to equilibrium
of a quantity as $\overline{\delta N}_1(t)$ and its possible
relation with local dynamical coefficients -- such as the Lyapunov 
exponent~\cite{schuster_01,beck_01,robledo_01} --
becomes thus particularly interesting at a fundamental level. In what
follows, our aim is to give a first numerical account of such a study,
which certainly deserves further insight in the future.

Taking for simplicity $L=2$ (in natural dimensionless units), 
Eqs.~\eqref{eq_wiener} are now replaced by the
logistic-map iterations
\begin{equation}
x_i(t+1)=1-\mu_{\rm lm}\;x_i(t)^2
\qquad(x_i\in[-1,1]).
\label{eq_logistic}
\end{equation}
For each particle, the iterates tend to an attractor 
whose characteristics depend on the value of the control parameter  
 $0\leq\mu_{\rm lm}\leq2$. 
Specifically, with $\mu_{\rm lm}=2$ the attractor's dynamics is fully
chaotic (positive Lyapunov exponent)~\cite{schuster_01,beck_01,robledo_01}.
Although the dynamics in Eq.~\eqref{eq_logistic} is now deterministic,
a positive exponential divergence of two initially close initial
conditions implies that any randomness in the definition of the
initial coordinates results in a random behavior for $N(x,t)$.
The latter is again defined as in Eq.~\eqref{eq_dist},
with the Lagrangian coordinates $x_i(t)$ substituted now by $N_0$
independent copies of the logistic map's coordinates, 
each evolving through Eq.~\eqref{eq_logistic}.

\begin{figure}
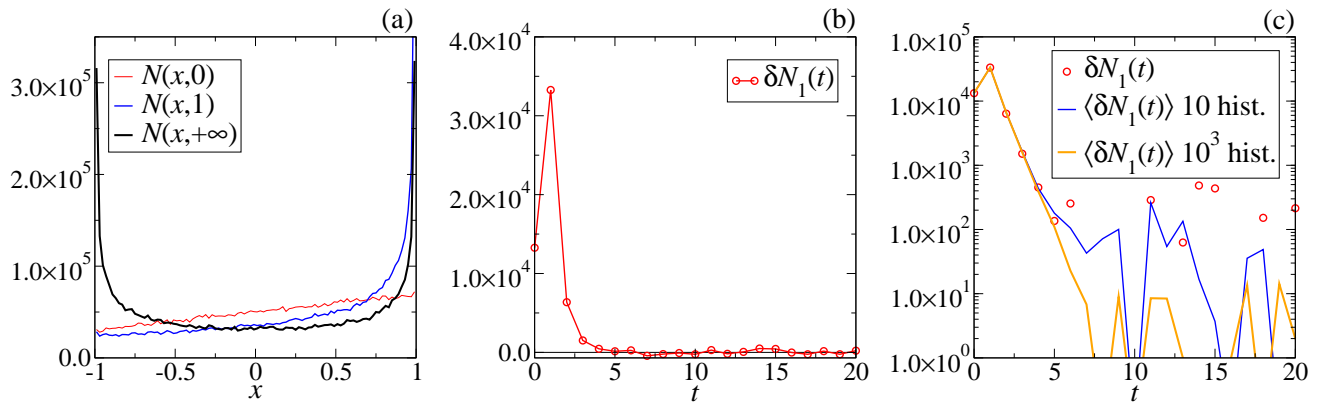

\begin{center}
\includegraphics[width=0.32\textwidth]{logistic_mu2_weight.eps}
\includegraphics[width=0.32\textwidth]{logistic_mu2_moment.eps}
\includegraphics[width=0.32\textwidth]{logistic_mu2_moment_log.eps}
\caption{ Chaotic conduction.
  The same numerical analysis displayed in Fig~\ref{fig_wiener} is
  performed replacing 
  Eqs.~\eqref{eq_wiener} with 
  Eqs.~\eqref{eq_logistic} and $\mu_{\rm lm}=2$.
  (a): Plots of the distribution $N(x,t)$ at different time.
  (b): Numerical evaluation of $\delta N_1(t)$ (the line is a guide to
  the eye).
  (c) The numerical evaluation of $\delta N_1(t)$ for a single history (circles) is
  compared with averages over many histories (lines).
}
\label{fig_logistic_mu2}
\end{center}
\end{figure}

For the sake of simplicity, in Fig.~\ref{fig_logistic_mu2} we
consider the same (linear) $N(x,0)$ used for the Wiener process in the previous
Section. With a single iteration, the chaotic map quickly drives this
initial distribution close to the equilibrium one, 
$N_{\rm eq}(x)=\lim_{t\to+\infty}\overline{N}(x,t)$; the latter is in this case
$x$-dependent with a characteristic ``U'' shape (see
Fig.~\ref{fig_logistic_mu2}a)~\cite{schuster_01,beck_01}. 
In parallel, apart from the initial
value, the time evolution of $\delta N_1(t)$ reported in
Fig.~\ref{fig_logistic_mu2}b displays features of a monotonic
decay. The log-linear plot of
Fig.~\ref{fig_logistic_mu2}c, where averages are also taken over
different histories sharing the same $N(x,0)$, provides evidence of an exponential 
decay~\cite{footnote_1}.

\section{Conduction at the onset of chaos}
At the chaos threshold
$\mu_{\rm lm}=1.401155189092\ldots$ 
(the period-doubling accumulation point~\cite{schuster_01,beck_01,robledo_01}) 
the Lyapunov exponent collapses to zero and to get sensible
information about the microscopic dynamics one is forced to consider
an infinite series of specific time-subsequences
and to replace exponential divergence (and convergence) 
with a spectrum of power-laws. 
Correspondingly, the
Lyapunov exponent must be substituted by an infinite series of 
generalized ones~\cite{robledo_01,baldovin_01,mayoral_01,fuentes_01}.

\begin{figure}
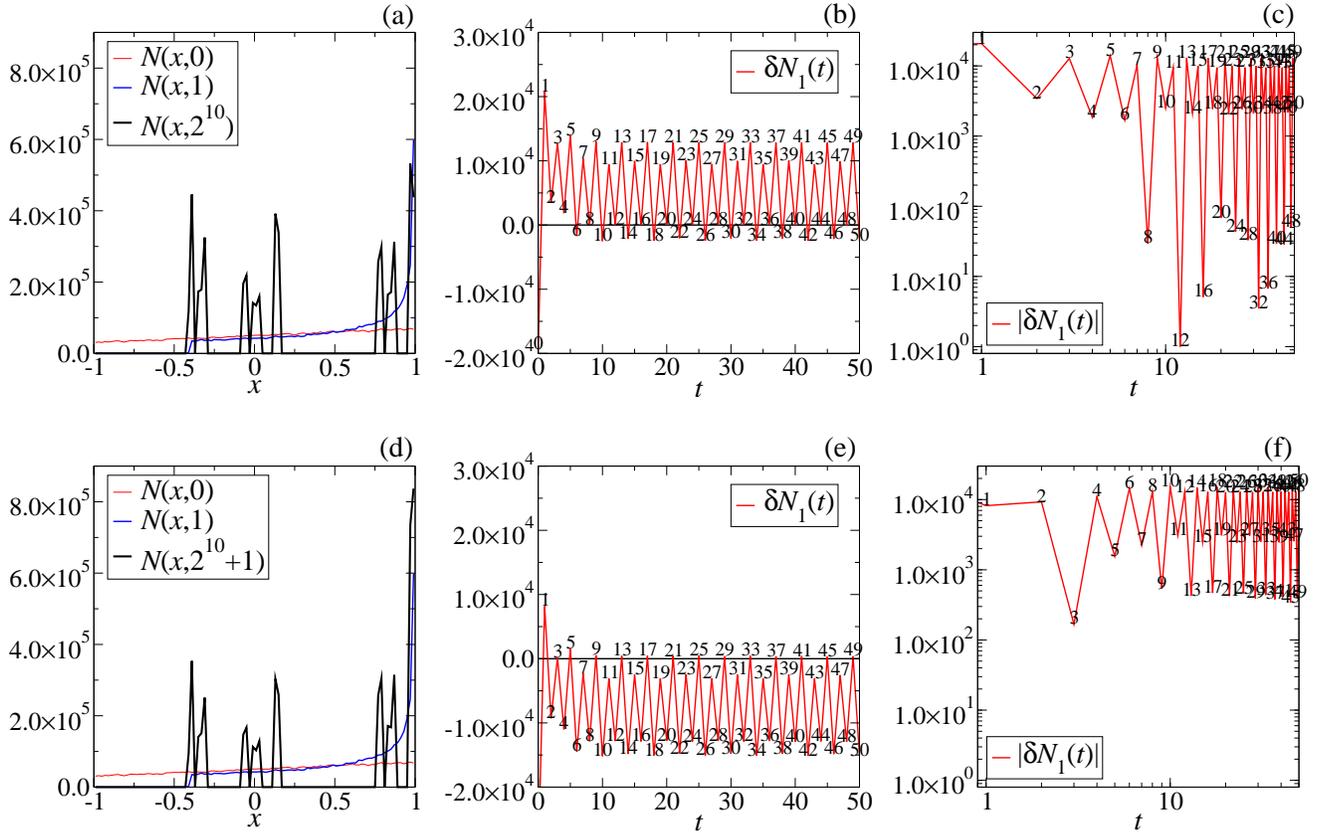

\begin{center}
\includegraphics[width=0.32\textwidth]{logistic_muc_weight.eps}
\includegraphics[width=0.32\textwidth]{logistic_muc_moment.eps}
\includegraphics[width=0.32\textwidth]{logistic_muc_moment_log.eps}
\includegraphics[width=0.32\textwidth]{logistic_muc_weight_odd.eps}
\includegraphics[width=0.32\textwidth]{logistic_muc_moment_odd.eps}
\includegraphics[width=0.32\textwidth]{logistic_muc_moment_odd_log.eps}
\caption{Conduction at the onset of chaos.
  The numerical analysis is performed with 
  Eqs.~\eqref{eq_logistic} and $\mu_{\rm lm}=1.401155189092\ldots$.
  (a,d): Plots of the distribution $N(x,t)$ at different time.
  (b,e): Numerical evaluation of $\delta N_1(t)$ defined through
  Eq.~\eqref{eq_delta_logistic_muc} with $\overline{t}=2^{10}$ in (b)
  and $\overline{t}=2^{10}+1$ in (e)  (the line is a guide to
  the eye).
  (c,f) Same as (b,e), in log-log scale.
}
\label{fig_logistic_muc}
\end{center}
\end{figure}

Fig.~\ref{fig_logistic_muc} displays the numerical analysis performed
starting with the same (linear) $N(x,0)$ of the previous cases, 
for the logistic map at the onset of chaos 
$\mu_{\rm lm}=1.401155189092\ldots$. 
As to be expected, results are now much more involved. 
In Fig.~\ref{fig_logistic_muc}a it is shown that 
the first iteration sets to zero $N(x,1)$ for $x$ smaller then about
$x=-0.5$, in correspondence of the (first) gap formation~\cite{schuster_01,beck_01,robledo_01,fuentes_01}.  
Then,  the 
long-time distribution $N(x,t\gg1)$ reflects the
multi-fractal properties of the attractor at the edge of chaos, being
characterized by many spikes and gaps. 
Indeed, initially trajectories are spread
out in the interval $[-1,1]$ and, except for the few that are initiated
inside the multifractal attractor, they get there via a sequence of gap
formations~\cite{schuster_01,beck_01,robledo_01,fuentes_01}. 
In practice most of them get into the attractor fairly soon,
so, after a few iterations, the first moment is built from positions of the
attractor, which are formed by bands (with inner gaps) separated
by (main) gaps (see, e.g., Fig. 2 in Ref.~\cite{fuentes_01}).
This implies $N(x,t)\simeq0$ for $x$ not in the attractor, if $t$ is sufficiently large.

A careful comparison of Fig.~\ref{fig_logistic_muc}a with
Fig.~\ref{fig_logistic_muc}d also reveals that in the present case it
is not sufficient to take the long-time limit of $N(x,t)$ to get the
(invariant) equilibrium distribution, since sensible differences can
be appreciated, e.g., between $N(x,2^{10})$ and $N(x,2^{10}+1)$.
With this in mind, our numerical study proceeds defining
\begin{equation}
\delta N_1(t)
\equiv A\;\int \mathrm{d} x\;x\;\left[N(x,t)-N(x,\overline{t})\right],
\label{eq_delta_logistic_muc}
\end{equation}
with $\overline{t}=2^{10}$ (Figs.~\ref{fig_logistic_muc}b
and~\ref{fig_logistic_muc}c) 
and $\overline{t}=2^{10}+1$ (Figs.~\ref{fig_logistic_muc}e
and~\ref{fig_logistic_muc}f).
Specifically, in Figs.~\ref{fig_logistic_muc}b
and~\ref{fig_logistic_muc}e we see that in both cases 
$\delta N_1(t)$ has an oscillating behavior in which even and odd time
iterations are well separated. 
For further specific time subsequences, 
the log-log plots of $|\delta N_1(t)|$ in Figs.~\ref{fig_logistic_muc}c
and~\ref{fig_logistic_muc}f may recall power-law
behaviors reminiscent to those characterizing the generalized Lyapunov
spectrum~\cite{baldovin_01,mayoral_01}, although deeper insight is certainly needed
to get definite results.

\section{Conclusions and perspectives}
In this paper we studied the relaxation process of a nonequilibrium
fluctuation in a context in which the particles'
dynamics is described by logistic map iterations. 
This allowed us to explore whether some of the features of the classic
Onsager's regression description generalize to non-local,
non-reversible microscopic dynamics. 

After a general discussion of
conductive processes in which simple thermodynamic observables have been
introduced, the conventional example of Browninan particles has been
analytically and numerically worked out. In this way, contact has been
established among the underlying dynamics
and system's geometry, and the thermodynamic behavior. 

Substituting the Brownian dynamics with logistic map iterations,
we numerically analyzed the same relaxation process.
While evidence of a monotonic relaxation has been found
when the control parameter $\mu_{\rm lm}$ is tuned to
chaoticity~\cite{footnote_2}, at the
onset of chaos a much more involved dynamical picture emerges, which may be
rationalized in terms of specific
time subsequences. Clarification of the latter result demands for a
better construction of the invariant (equilibrium) measure than the one
obtained by simply taking the long-time limit of the particles'
distribution. 
More generally, a statistical mechanics approach~\cite{diaz_01} linking the
microscopic dynamics (e.g., in terms of the Lyapunov or generalized
Lyapunov exponents) to the observed nonequilibrium thermodynamics
is an intriguing open question.  

\section*{Acknowledgments}
A. D\'{\i}az-Ruelas and A. Robledo are acknowledged for important
discussions and remarks.

\end{document}